\RequirePackage{fix-cm}
\documentclass[smallextended]{svjour3}
\usepackage{graphicx}
\graphicspath{{figures/}}

\usepackage{mathptmx}      

\newcommand{\PANDA}{$\overline{\mbox{P}}${\sc anda}}

\journalname{EXA2011 Conference Proceedings}
\begin{document}

\title{Hypernuclear Physics at
  $\overline{\mbox{P}}$ANDA\thanks{Supported by the Research
    Infrastructure Integrating Activity ``Study of Strongly
    Interacting Matter'' HadronPhysics2 (SPHERE) under the 7th
    Framework Programme of EU. We acknowledge financial support from
    the Bundesministerium f\"ur Bildung und Forschung (bmb+f) under
    contract number 06MZ9182.} }

\subtitle{Experimental Challenges}

\author{P.~Achenbach \and S.~Bleser \and J.~Pochodzalla \and
        A.~Sanchez~Lorente \and M.~Steinen}
        
\institute{P. Achenbach \and S. Bleser \and J. Pochodzalla \and
  M. Steinen \at Institut f\"ur Kernphysik,
  Johannes Gutenberg-Universit\"at \\ J.J.-Becherweg 45, D-55099
  Mainz, Germany \\ \email{patrick@kph.uni-mainz.de} 
  \and A. Sanchez Lorente \at Helmholtz-Institut Mainz \\ D-55099
  Mainz, Germany}

\date{Received: 20 September 2011 / Accepted: date}

\maketitle

\begin{abstract}
  Hypernuclear research will be one of the main topics addressed by
  the \PANDA\ experiment at the planned Facility for Anti-proton and
  Ion Research FAIR at Darmstadt, Germany. A copious production of
  $\Xi$-hyperons at a dedicated internal target in the stored
  anti-proton beam is expected, which will enable the high-precision
  $\gamma$-spectroscopy of double strange systems for the first
  time. In addition to the general purpose \PANDA\ setup, the
  hypernuclear experiments require an active secondary target of
  silicon layers and absorber material as well as high purity
  germanium (HPGe) crystals as $\gamma$-detectors.  The design of the
  setup and the development of these detectors is progressing: a first
  HPGe crystal with a new electromechanical cooling system was
  prepared and the properties of a silicon strip detector as a prototype
  to be used in the secondary target were studied. Simultaneously to the
  hardware projects, detailed Monte Carlo simulations were performed
  to predict the yield of particle stable hypernuclei. With the help
  of the Monte Carlo a procedure for $\Lambda\Lambda$-hypernuclei
  identification by the detection and correlation of the weak decay
  pions was developed.

\end{abstract}

\section{Introduction}

The ``standard model'' is the fundamental theory which unites weak,
electromagnetic, and strong interactions. Strong processes are
formulated in Quantum Chromodynamics (QCD), the field theory for the
dynamics of quarks and gluons.  QCD has been thoroughly probed in
strong interactions at very high energies.  However, at the energy
scale of the nucleon mass hadrons are complex many-body systems. 
Even though they interact strongly, the description by the
fundamental QCD equations is complicated by the non-perturbative
nature of the theory.  The investigation of strange hadrons carrying
an additional flavour degree-of-freedom is essential for understanding
the low-energy regime of QCD.

A very interesting phenomenon in nuclear physics is the existence of
nuclei containing strange baryons. The lightest hyperons are stable
against strong and electromagnetic decays, and as they do not suffer
from Pauli blocking by other nucleons they can live long enough in the
nuclear cores to become bound. When a hyperon, specifically a
$\Lambda$-hyperon, replaces one of the nucleons in the nucleus, the
original nuclear structure changes to a system composed by the hyperon
and the core of the remaining nucleons.  The existence of double
strange nuclear systems like $\Xi^-$ or $\Lambda\Lambda$ hypernuclei
is directly linked to the strength of the attractive hyperon--nucleon
($YN$) and unknown hyperon--hyperon ($YY$) interactions.  Models for
the $YY$ interactions have been constructed from the expansion of the
$NN$ interaction and limited $YN$ scattering data. This approach needs 
to be validated against double hypernuclei binding energies and excitation
spectra.

The study of strange nuclear systems provides invaluable information
on both, on the structure of nuclei as many-body hadronic systems and
on strange baryons in the nuclear medium. Although single and double
$\Lambda$-hypernuclei were discovered many decades ago in cosmic ray
interactions studied by the emulsion technique, only few double
$\Lambda$-hypernuclear isotopes are presently known.  In particular,
$\Lambda\Lambda$-hypernuclei formed in anti-proton beams are the only
practical systems among all strange baryons for investigating the
strong nuclear interaction.

\section{Production and detection of hypernuclei at 
  $\overline{\mbox{P}}$ANDA}

The planned Facility for Anti-proton and Ion Research (FAIR) near
Darmstadt will include the High Energy Storage Ring (HESR) to store
anti-protons of several GeV$\!/c$ momentum in an intense and high
quality beam.  With a dedicated internal target in the storage ring a
copious production of $\Xi$-hyperons is expected which can be stopped
in dedicated absorbers to form bound states of $\Xi$ hypernuclei. The
latter can be used as a gateway to form $\Lambda\Lambda$ hypernuclei,
which will enable the high-precision $\gamma$-spectroscopy of double
strange systems for the first time~\cite{Pochodzalla2004,PANDA2009}.

The \PANDA\ experiment (AntiProton ANnihilations at DArmstadt) planned
at the HESR storage ring is a next-generation hadron physics
experiment. In addition to the general purpose \PANDA\ setup for
charged particle detection, the hypernuclear experiments require an
active secondary target of silicon layers and absorber material in
addition to high purity germanium (HPGe) crystals as
$\gamma$-detectors.  Some technical and practical aspects currently
being studied by the \PANDA\ hypernuclear groups are
\begin{enumerate}
  \item design and fabrication of the primary target,
  \item design and development of the secondary target, and
  \item design and operation of the HPGe $\gamma$-array.
\end{enumerate}
In the following we highlight some of the experimental challenges in
realizing the setup. The expected performance of the proposed
experiment with this setup was simulated with the help of a
micro-canonical decay model was redicting the yield of particle stable
double hypernuclei.

\section{The challenge for a nuclear internal storage ring target}

On a nuclear internal target low momentum $\Xi$ pairs can be produced
in $\mathrm{\overline{p}p} \to \Xi^- \overline{\Xi}^+$ or
$\mathrm{\overline{p}n} \to \Xi^- \overline{\Xi}^0$ reactions. The
advantage at HESR in the $\Xi$ production rate as compared to kaon
beam induced reactions is the fact that the anti-proton is stable and
can be retained in the storage ring.  The largest
$\mathrm{\overline{p}}$ production rate achievable at HESR is of the
order of $10^7\,\mathrm{\overline{p}}/$s with a maximum of
approximately $10^{11}$ stored anti-protons in the HESR
ring~\cite{PANDA2009}.  This allows a rather high luminosity even with
very thin primary targets, either with the standard hydrogen target of
the \PANDA\ experiment, or with a dedicated target for hypernuclear
spectroscopy.

The beam--target interactions will reduce the life-time of the stored
anti-proton beam. Consequently, a reasonable compromise between
$\mathrm{\overline{p}}$ beam preparation time and beam loss rate
during one HESR cycle needs to be found. Obviously, this compromise is
dependent on the target material and thickness. The possibility of
steering a low density region of the transverse beam profile over the
target along with the gradual consumption of anti-protons will be an
important feature.  These are conditions posing a real challenge for
the design of the nuclear internal target inside the storage ring.

At present, techniques for the manufacturing of $\mu$m-thin synthetic
diamond filaments cut from a membrane as internal target are being
explored~\cite{Ferro2011}. Synthetic diamonds can be grown from a
hydrocarbon gas mixture by chemical vapour deposition and can be
supported by a silicon ring structure matching the beam pipe.  The
thermal conductivity of synthetic diamond is very high that prevents
the materials from overheating. The lateral size of the synthetic
diamond filament could be of the order of 100\,$\mu$m or less. The
target could be designed with an empty region inside the support ring
providing space during beam preparation and for the high-density beam
center during the anti-proton storage. Such geometries can be made
very precisely by cutting the diamond membrane with fs-pulsed
high-powered lasers~\cite{Ferro2011}.

\section{The challenge for an active high-resolution $\Xi$-absorber 
  target}

The main purpose of the active high-resolution $\Xi$-absorber target
is the tracking and stopping of the produced cascade hyperons and
their decay products.  The active part of the secondary target will be
made from silicon strip sensors.  The slowing down of the $\Xi^-$
proceeds by energy-loss during the passage through additional absorber
layers. If decelerated to rest before decaying, the $\Xi^-$ can be
captured inside a nucleus, eventually releasing two $\Lambda$ hyperons
and forming a double hypernuclei. The geometry of the target is
essentially determined by the life-time of the hyperons and their
stopping time in silicon and the absorber materials.

In analogy to the germanium detectors array, the silicon detector has
to be able to operate in extreme conditions such as a large hadronic
environment since it is close to the interaction point.  Furthermore,
the material budget on the detector volume must be kept low.  The
feasibility of such a device has recently been studied in Mainz with a
single sided microstrip sensor with a strip pitch of 50\,$\mu$m and
2$\times$2\,cm$^2$ size. In addition, studies on ultra-thin cables
based on adhesive-less aluminium-polyimide foiled dielectrics, the
128-channel APV-25-S1 front-end chip~\cite{DallaBetta2001}, the
modified electronics board, and the mechanics of the target are being
performed.

\section{The challenge for a compact $\gamma$-array inside the 
  $\overline{\mbox{P}}$ANDA setup}

\begin{figure}
  \centering
  \includegraphics[width=0.7\columnwidth]{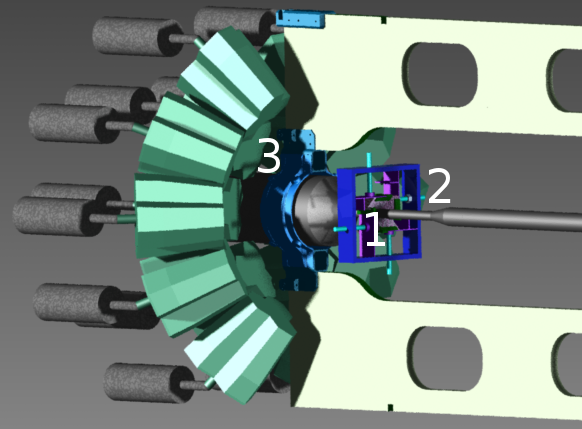}
  \caption{Scheme of the hypernuclear setup inside \PANDA. In the
    backward region of the spectrometer a dedicated synthetic diamond
    target will be installed inside a beam pipe of reduced diameter
    (1). A mechanical support structure surrounds this target and
    holds the active $\Xi$-absorber target (2). A compact
    $\gamma$-array with maximized solid-angle acceptance will be used
    to detect radiative de-excitations (3).}
  \label{fig:setup}
\end{figure}

High purity germanium (HPGe) detectors are key instruments in nuclear
structure physics for detecting the radiative de-excitation of excited
nuclei. Typically, the crystals are cooled with liquid nitrogen and
operated in the temperature range of 77--115\,K.  The requirement of
minimal distance of the secondary target from the interaction point
combined with the necessary support structures nested inside the
\PANDA\ target spectrometer leaves very restricted space for the
installations of the hypernuclear physics setup.  The situation is
displayed in Fig.~\ref{fig:setup}, showing the anti-proton beam pipe
surrounded by the targets and an $\gamma$-array of 15 n-type HPGe
triple cluster detectors.

\begin{figure}
  \centering
  \includegraphics[width=0.7\columnwidth]{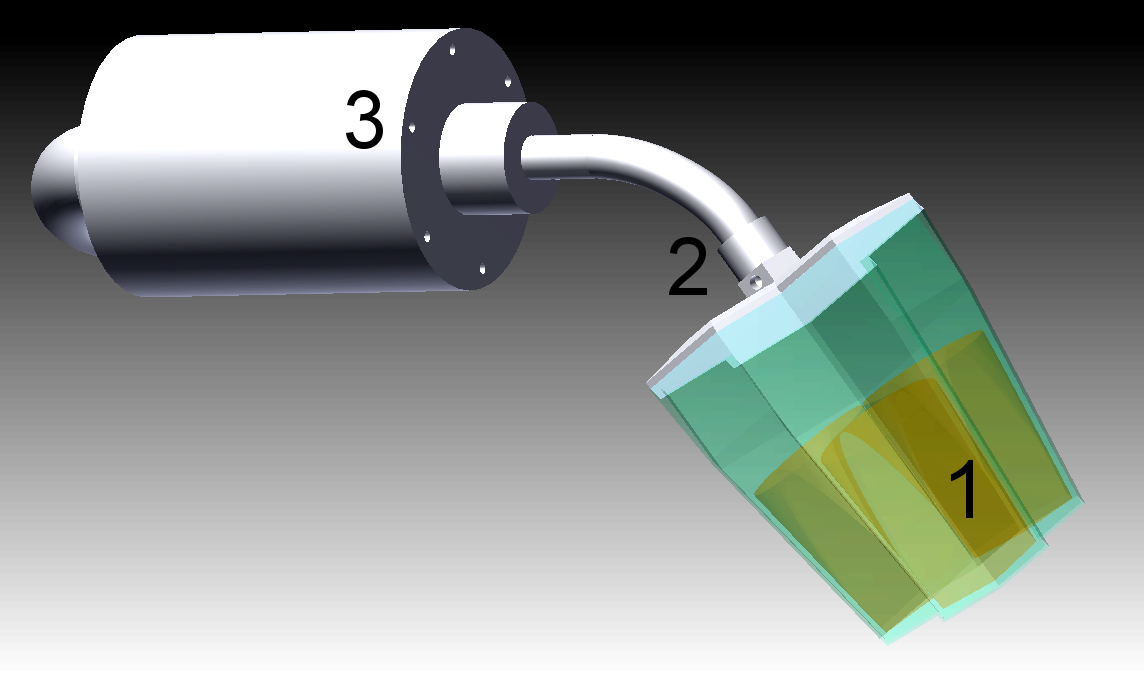}
  \caption{Drawing of one HPGe detector system assembled to an
    X-Cooler~II device. Three encapsulated coaxial HPGe crystals (1)
    are arranged in one capsule. The flexible section of the thick
    cold finger (2) enables the placement of the cluster at the
    restricted space inside the \PANDA\ spectrometer. The X-Cooler~II
    (3) replaces the standard liquid nitrogen cooling devices.}
  \label{fig:HPGe}
\end{figure}

For an effective integration of the array into the
\PANDA\ spectrometer an electromechanical cooling device will be used
instead of liquid nitrogen with its bulky dewars. Composite detectors
made of three large volume encapsulated Ge crystals and cooled by the
electromechanical cooling device X-Cooler~II by ORTEC are being
considered. Fig.~\ref{fig:HPGe} shows the drawing of a triple HPGe
cluster assembled to an X-Cooler~II device.  The individual Ge
crystals are sealed in an aluminium can and installed in a common
vacuum cryostat.  An intermediate thermal shield may be applied in
order to act as a heat reflector thus reducing the heating of the
encapsulated Ge crystals.

\begin{figure}
  \centering
  \includegraphics[width=0.7\columnwidth]{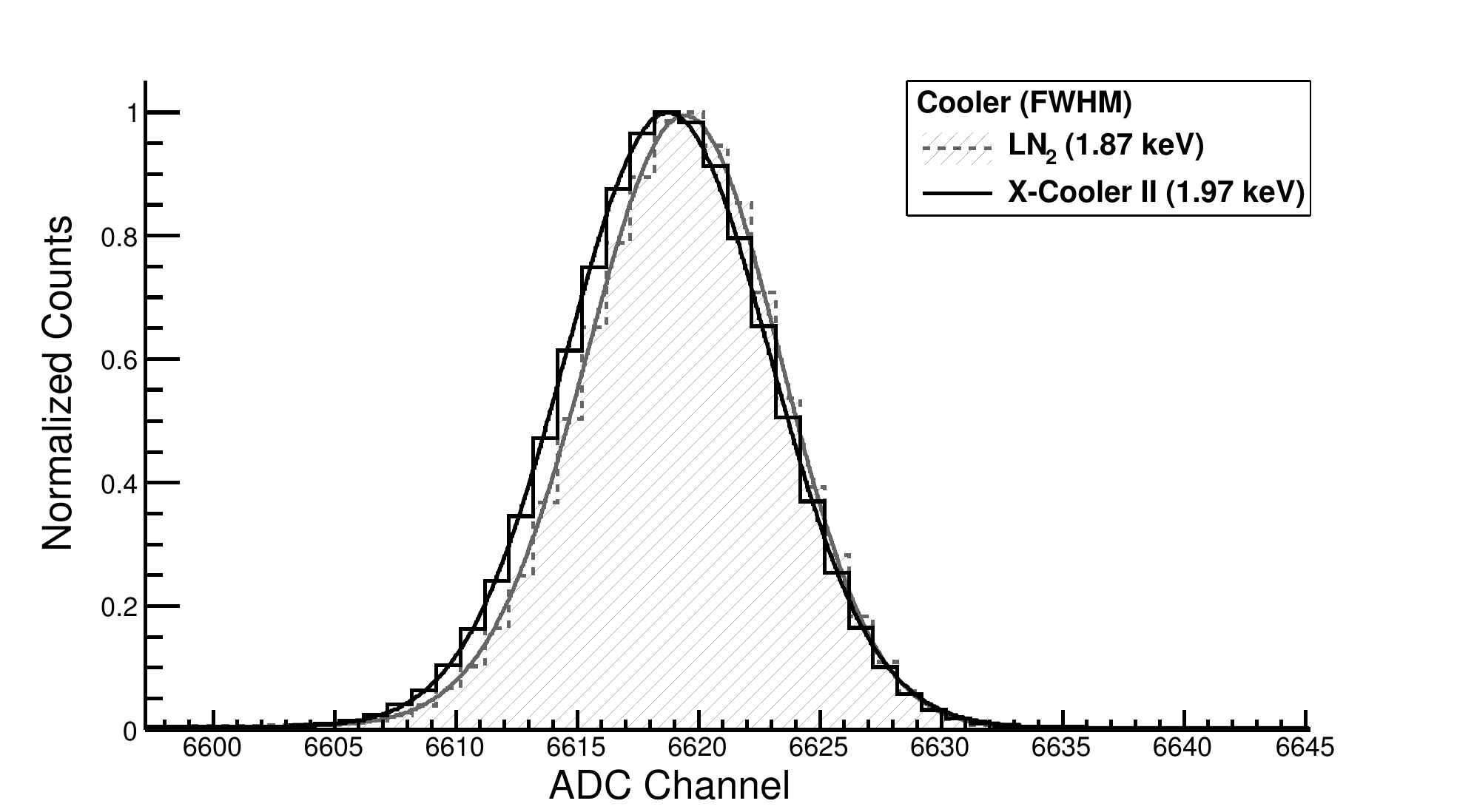}
  \caption{Measured energy spectra of the 1.332\,keV line of a
    $^{60}$Co calibration source taken with two different cooling
    devices.  For the dashed spectrum with a line width of FWHM =
    1.97\,keV the HPGe crystal was cooled electromechanically, for the
    solid spectrum with a line width of FWHM = 1.87\,keV a liquid
    nitrogen cooling system was used~\cite{SteinenDipl}.}
  \label{fig:Co60}
\end{figure}

The energy resolution of such a system with electromechanical cooling
has been determined with a standard $^{60}$Co calibration source and a
line width of 1.97\,keV for the 1.332\,keV $\gamma$-line was found,
see Fig.~\ref{fig:Co60}. In comparison to a cooling device based on
liquid nitrogen where a line width of 1.86\,keV was measured, the
electromechanical cooling seems to have no negative impact on the
performance~\cite{SteinenDipl}.

Another major challenge at \PANDA\ is the operation of the germanium
detectors close to a strong magnetic field over long periods. It was
demonstrated that a good energy resolution can be preserved up to
1\,T~\cite{Lorente2007}.

\section{Performance of the proposed hypernuclear experiment}

The production of excited states in $\Lambda\Lambda$ hypernuclei was
studied following the micro-canonical break-up of an initially excited
double hypernucleus created by the absorption and conversion of a
stopped $\Xi^-$ hyperon~\cite{Lorente2008,Lorente2011}. In these
calculations the formation of excited states dominates. Furthermore,
different double hypernuclei isotopes which depend on the initial
target nuclei are formed. Thus, the ability to assign the observable
$\gamma$-transitions in a unique way to a specific double
$\Lambda\Lambda$ hypernucleus seems possible.

The non-mesonic and mesonic decays of the light hypernuclei to be
studied in the initial phase of the planned experiments are of similar
magnitude. The analysis strategy will make use of the two-body pionic
decays, where the mono-energetic pions will leave a unique signature
in the secondary target. In the case of two sequential mesonic weak
decays of the double hypernuclei, the momenta of the two pions are
strongly correlated. Thus, a coincidence measurement in the active
$\Xi$-absorber target will provide an effective method to tag the
production of a double hypernucleus.

\section{Conclusions}

At the \PANDA\ experiment at FAIR it will become possible to explore
the level scheme of low-mass isotopes of double hypernuclei for the
first time. The spectroscopic information will be obtained via
$\gamma$-ray detection using an array of 15 n-type HPGe triple cluster
detectors located near a dedicated arrangement of targets: a primary
diamond target at the entrance to the central tracking detector of
\PANDA, and a small secondary active target composed of silicon
detectors and absorbers to decelerate and stop $\Xi$-hyperons and to
identify the weak decay products.

\end{document}